\documentclass[aps,
prb,
twocolumn,
groupedaddress,
amsmath,
amssymb,
]{revtex4-2}

\usepackage{graphicx}
\usepackage{dcolumn}
\usepackage{bm}
\usepackage{comment}

\begin{document}

\hyphenation{nanotube}
\hyphenation{nanotubes}


\title{
Spin-orbit torque driven motion of chiral domain walls induced by radial magnetization in nanotube geometry
}


\author{Nobuyuki Umetsu}
\affiliation{Frontier Technology R\&D institute, Kioxia Corporation, 3-13-1, Moriya-cho, Kanagawa-ku, Yokohama, 221-0022, Japan.}
\author{Hiroki Tokuhira}
\affiliation{Frontier Technology R\&D institute, Kioxia Corporation, 3-13-1, Moriya-cho, Kanagawa-ku, Yokohama, 221-0022, Japan.}
\author{Michael Quinsat}
\affiliation{Frontier Technology R\&D institute, Kioxia Corporation, 3-13-1, Moriya-cho, Kanagawa-ku, Yokohama, 221-0022, Japan.}
\author{Hideto Horii}
\affiliation{Frontier Technology R\&D institute, Kioxia Corporation, 3-13-1, Moriya-cho, Kanagawa-ku, Yokohama, 221-0022, Japan.}
\author{Tsuyoshi Kondo}
\affiliation{Frontier Technology R\&D institute, Kioxia Corporation, 3-13-1, Moriya-cho, Kanagawa-ku, Yokohama, 221-0022, Japan.}
\author{Masaki Kado}
\affiliation{Frontier Technology R\&D institute, Kioxia Corporation, 3-13-1, Moriya-cho, Kanagawa-ku, Yokohama, 221-0022, Japan.}

\date{\today}

\begin{abstract}
We theoretically investigate chiral domain walls (DWs)  formed in radially magnetized nanotubes composed of ultrathin layers with perpendicular magnetic anisotropy (PMA). Unlike those with in-plane magnetic anisotropy, the stable configurations of DWs in PMA nanotubes are influenced not only by exchange interactions but also by magnetostatic interactions induced by the radial component of magnetization. Particularly, the magnetostatic interactions lead to Dzyaloshinskii-Moriya interaction (DMI)-like effects that stabilize chiral N\'{e}el-type DWs.  We derive expressions for the effective magnetic fields acting on DWs within PMA nanotubes and quantify spin-orbit torque (SOT) driven DW motion using an analytical one-dimensional model, which is validated by micromagnetic simulations. Our results show that the DMI-like field due to magnetostatic interactions can be as significant as the contribution of material-induced DMI in nanotubes with diameters below $100\,$nm. This implies that the direction and speed of DW motion in the PMA nanotubes could differ from those observed in flat nanoribbons composed of the same material. Furthermore, we demonstrate that DW velocity can be effectively controlled by adjusting the tube diameter and exchange stiffness constant of the magnetic layer, rather than relying solely on material-induced DMI. These insights are expected to greatly expand the potential applications of PMA nanotube-based DW devices.
\end{abstract}


\maketitle


\section{Introduction}

Recent studies on the behavior of magnetic textures in curved magnetic layers have garnered significant attention
\cite{Streubel_2016, Streubel_2021,Makarov_2022}.
It has been theoretically predicted that the curvature and torsion of magnetic films affect the dynamics of magnetic domain walls (DWs)
\cite{Yershov_2015, Yershov_2016, Moreno_2017, Yershov_2018}, with some predictions corroborated by experimental evidence
\cite{Krugner_2007, Garg_2017, Volkov_2019, Skoric_2022, Farinha_2025}.
Understanding these phenomena could enable the storage and manipulation of DWs within three-dimensional (3D) structures
\cite{Makarov_2022, Pacheco_2017},
paving the way for high-density data storage applications
\cite{Parkin_2008, Parkin_2015}.

In this study, we focus on nanotubes as a promising structure for future 3D magnetic DW memory (DWM) \cite{Parkin_2008, Parkin_2015, Stano_2018, Quinsat_2025}.
Magnetic nanotubes and nanocylinders can be fabricated using methods such as
rolling up of thin films \cite{Prinz_2000, Schmidt_2001, Streubel_2014, Streubel_2015},
electroplating \cite{Borissov_2009, Biziere_2013, Col_2014, Ivanov_2016, Bran_2016, Bran_2017, Bran_2018, Stano_2018, Ruiz_2018, Proenca_2021,Donnely_2022, Huang_2023, Tiwari_2023}, and atomic layer deposition (ALD) \cite{Ruffer_2012, Weber_2012, Buchter_2013,Jussila_2023}.
Compared to nanocylinders, nanotubes have the advantage of supplying higher current densities for the same applied current due to their smaller cross-sectional area, which is crucial for generating sufficient torque for DW motion.
Although ALD techniques for magnetic structures are still in their early stages, recent observations of current-induced DW motion in ALD-Co layers \cite{Kado_2023}
lead us to believe that magnetic nanotubes utilizing ALD techniques represent a promising candidate for the essential components of 3D-DWM.

Previous theoretical research on the properties of DWs in nanotubes has primarily focused on systems with in-plane magnetic anisotropy (IMA)
\cite{Landeros_2010, Yan_2011, Yan_2012, Otalora_2012_APL, Otalora_2012_JPCM, Otalora_2013, Goussev_2016, Pylypovskyi_2016, Depassier_2019, Yershov_2020, Almonacid_2020, Hurst_2021}.
These studies have predominantly considered DW motion driven by conventional spin-transfer torque (STT) and {\O}rsted field \cite{Otalora_2012_JPCM, Hurst_2021}.
In contrast, this study aims to establish a theoretical framework for DW dynamics in radially magnetized nanotubes composed of thin films with perpendicular magnetic anisotropy (PMA) \cite{Streubel_2015, Bao_2022}.
Such systems are expected to utilize spin-orbit torque (SOT)
\cite{Thiaville_2012, Khvalkovskiy_2013, Martinez_2013}, potentially enabling more efficient DW motion compared to traditional driving forces.
Therefore, PMA nanotube-based DW shift registers are considered highly promising candidates for low power consumption data storage applications.

Generally, geometrical effects are known to arise from exchange interactions
\cite{Sheka_2015, Gaididei_2017}; however, it has been predicted that geometrical effects due to magnetostatic interactions can also occur, depending on the curvature characteristics and the easy axis of magnetization
\cite{Sheka_2020}.
Previous studies indicate that such magnetostatic interactions in tubular structures induce geometrical effects only in systems exhibiting a magnetization component along the radial axis of the tube \cite{Sheka_2020, Otalora_2016},
leading to energy contributions similar to the Dzyaloshinskii-Moriya interaction (DMI)
\cite{Dzyaloshinsky_1958, Moriya_1960}.
This suggests the potential induction of chirality in DWs within tubular structures.
The magnitude of DMI is crucial for the efficiency of SOT-driven DW motion
\cite{Thiaville_2012},
making it essential to theoretically predict the impact of these geometrical effects, to evaluate the potential applications of PMA nanotubes.

The purpose of this study is to elucidate the characteristics of chiral DWs in PMA nanotubes and quantitatively predict the influence of geometrical effects on SOT-driven DW motion. Specifically, we derive expressions for the effective magnetic fields acting on DWs due to magnetostatic interactions and formulate an analytical model for SOT-driven DW motion.
The validity of this model is confirmed through micromagnetic (\textmu M) simulations.

The remainder of this paper is organized as follows.
In Section II, we describe the model of PMA nanotubes.
Section III presents the expressions for effective magnetic fields due to geometrical effects in PMA nanotubes and discusses the characteristics of these fields and the stability of DWs as a function of nanotube diameter.
In Section IV, we present the results of SOT-driven DW motion calculations, comparing the outcomes from the analytical one-dimensional model (1DM) 
\cite{Malozemoff_1979, Tatara_2008, Thiaville_2012, Martinez_2014, Rizinggard_2017} and \textmu M simulations \cite{Martinez_2014}.
We also discuss the comparative shift characteristics of nanotubes and flat nanoribbons, as well as the impact of variations in tube diameter and magnetic properties on these characteristics.
Finally, Section V summarizes the key findings and implications of this study.

\section{Model of radially magnetized nanotube}

\begin{figure}[htbp]
\centering{}\includegraphics[scale=0.4]{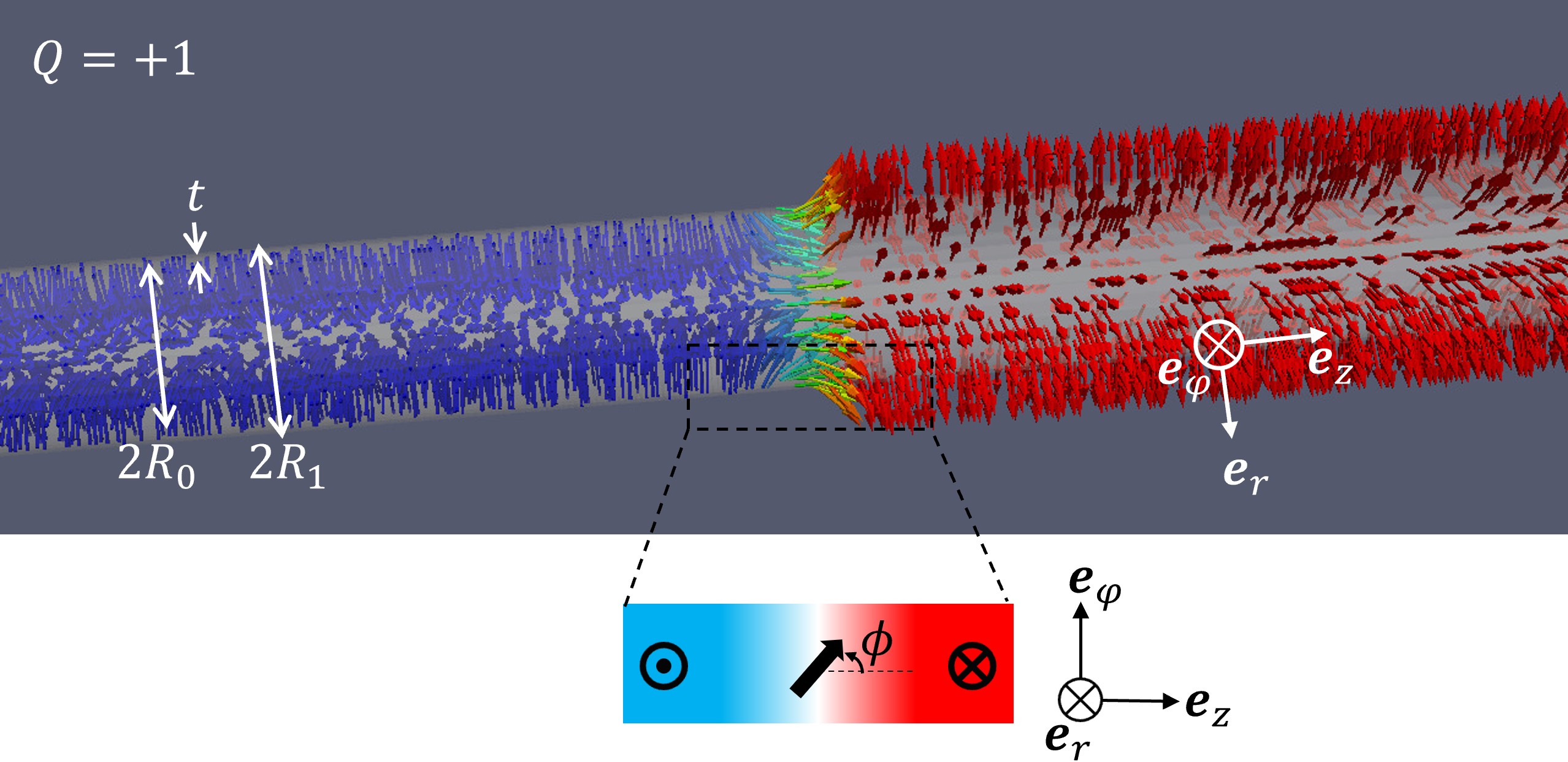}\caption{
Schematic illustration of a radially magnetized nanotube containing a DW.
The magnetic structure for an up-down DW configuration ($Q=+1$) is presented.
The enlarged diagram shows the magnetization configuration of the DW as seen from the inside of the tube.
\protect\label{fig:model}}
\end{figure}

We consider a magnetic nanotube that extends in the $z$ direction, with inner and outer radii denoted as $R_{0}$ and $R_{1}$, respectively (Fig. \ref{fig:model}).
The thickness of the nanotube is defined as $t=R_{1}-R_{0}$, with the assumption that $t\ll R_{0}$.
We assume that the nanotube is of infinite length, allowing us to neglect any end effects. 
Our study focuses on DWs in nanotubes composed of heavy metal (HM)/ferromagnet (FM) bilayers,
where the interfacial DMI induces SOT-driven DW motion \cite{Moore_2008, Kim_2010, Miron_2011, Haazen_2013, Emori_2013, Koyama_2013, Torrejon_2014, Emori_2014, Ryu_2014, Torrejon_2016, Lau_2019, Umetsu_2025}.
The normalized magnetization vector is expressed using the basis vectors of
the cylindrical coordinate system: $\bm{m}=m_{r}\bm{e}_{r}+m_{\varphi}\bm{e}_{\varphi}+m_{z}\bm{e}_{z}$,
where $\bm{e}_{r}$, $\bm{e}_{\varphi}$, and $\bm{e}_{z}$ are the radial, azimuthal, and axial directions, respectively.

The total energy of our model comprises contributions from the PMA, DMI, exchange interactions, and magnetostatic interactions (demagnetization) terms.
We assume nanotubes made of typical magnetic materials consisting of 3$d$ transition metal alloys,
characterized by the following parameters:
the saturation magnetization $M_{{\rm s}}$ ($=10^{6}\,{\rm A/m}$),
the exchange stiffness constant $A$,
the magnetic anisotropy constant perpendicular to the magnetic layer $K_{{\rm u}}$, and the DMI constant $D$.
The sign of $D$ is defined under the assumption that the outer side of the tube is the substrate side.
When $D > 0$ ($D < 0$), a right (left)-handed  N\'{e}el-type DW is favored.
To focus on the geometrical effects induced by magnetization configurations constrained by the nanotube shape and the magnetic anisotropy,
the modulation of electronic states by the curvature of the nanotube \cite{Edstrom_2022} are neglected in our model ; instead, we assume these magnetic parameters to be constants.

In PMA nanotubes, the out-of-plane direction perpendicular to the film surface coincides with the radial direction of the tube, thus the PMA energy density can be expressed as $\varepsilon_{{\rm PMA}} =-K_{{\rm u}}m_{r}^{2}$.
Additionally, the DMI and exchange energy densities can be respectively expressed as
$\varepsilon_{{\rm DMI}} =D\left[m_{r}\frac{\partial}{\partial z}m_{z}-m_{z}\frac{\partial}{\partial z}m_{r} + \frac{1}{r}\left(m_{r}\frac{\partial}{\partial \varphi}m_{\varphi}-m_{\varphi}\frac{\partial}{\partial \varphi}m_{r} \right) \right]$
and
$\varepsilon_{{\rm exc}} =A\sum_{i=x,y,z}\left[\left(\frac{\partial m_{i}}{\partial r}\right)^{2}+\frac{1}{r^{2}}\left(\frac{\partial m_{i}}{\partial\varphi}\right)^{2}+\left(\frac{\partial m_{i}}{\partial z}\right)^{2}\right],$
where $m_{x}=m_{r}\cos\varphi-m_{\varphi}\sin\varphi$ and $m_{y}=m_{r}\sin\varphi+m_{\varphi}\cos\varphi$.
The energy of  magnetistatic interaction is expressed as:
\begin{equation}
\begin{aligned}
E_{{\rm dem}}&=E_{{\rm dem,bb}}+E_{{\rm dem,ss}}+E_{{\rm dem,bs}},\\
E_{{\rm dem,bb}} & =\frac{1}{2}\frac{\mu_0}{4\pi}\int{\rm d}^{3}\bm{x}\int{\rm d}^{3}\bm{x}'\frac{\rho_{{\rm m}}\left(\bm{x}\right)\rho_{{\rm m}}\left(\bm{x}'\right)}{\left|\bm{x}-\bm{x}'\right|},\\
E_{{\rm dem,ss}} & =\frac{1}{2}\frac{\mu_0}{4\pi}\int{\rm d}^{2}\bm{x}\int{\rm d}^{2}\bm{x}'\frac{\sigma_{{\rm m}}\left(\bm{x}\right)\sigma_{{\rm m}}\left(\bm{x}'\right)}{\left|\bm{x}-\bm{x}'\right|},\\
E_{{\rm dem,bs}} & =\frac{\mu_0}{4\pi}\int{\rm d}^{3}\bm{x}\int{\rm d}^{2}\bm{x}'\frac{\rho_{{\rm m}}\left(\bm{x}\right)\sigma_{{\rm m}}\left(\bm{x}'\right)}{\left|\bm{x}-\bm{x}'\right|}, \label{eq:Edem}
\end{aligned}
\end{equation}
where $\mu_0$ is the permeability of
free space, $\sigma_{{\rm m}}\left(\bm{x}\right)=M_{\rm s}\bm{m}\left(\bm{x}\right)\cdot\bm{n}_{\perp}\left(\bm{x}\right) $ is the surface magnetic charge density, with $\bm{n}_{\perp}\left(\bm{x}\right)$ being the normal vector of the tube surface,
and $\rho_{{\rm m}}\left(\bm{x}\right)=-M_{\rm s}\nabla\cdot\bm{m}\left(\bm{x}\right)$ is the volume magnetic charge density \cite{Sun_2014}.

\begin{figure}[htbp]
\centering{}\includegraphics[scale=0.68]{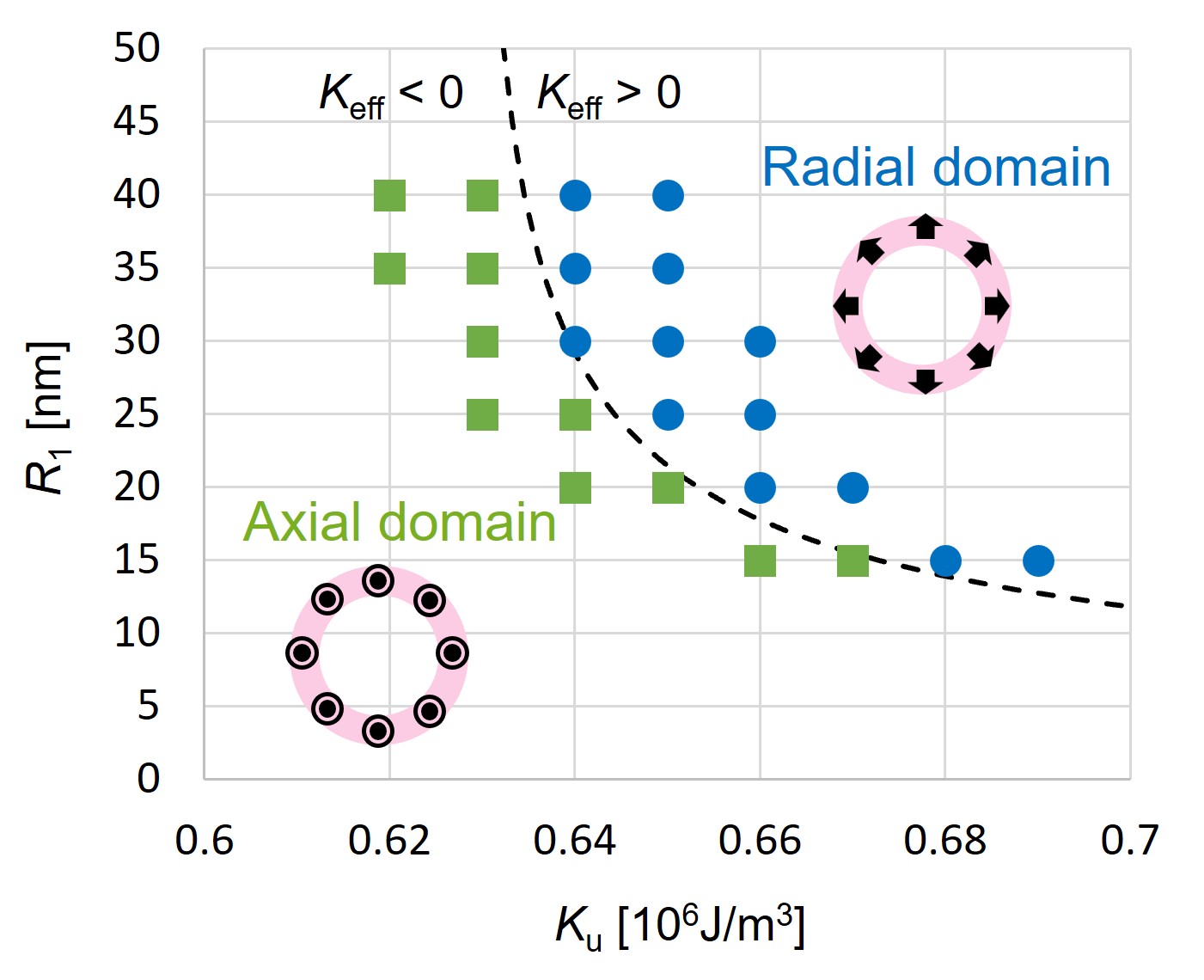}
\caption{\protect\label{fig:stable_states}
Phase diagram of magnetic domain structures in the PMA nanotube.
The dashed line represents the boundary of stable states as determined by Eq. (\ref{eq:Keff}).
The markers indicate conditions validated through  \textmu M simulations.
The following parameters are used;
$A=10^{-11}\,{\rm J/m}$ and  $t=1\,{\rm nm}$.
}
\end{figure}

Assuming the  demagnetizing factor in the radial direction is equal to 1 due to $t\ll R_0$,
and comparing the energy density of the uniform radial magnetization structure with that of the uniform axial magnetization structure,
the effective magnetic anisotropy constant for PMA layers of nanotubes is given by
\begin{equation}
K_{{\rm eff}}= K_{{\rm u}}-\frac{\mu_0}{2} M_{{\rm s}}^{2}
+\frac{A}{\left(R_1 - t/2\right)t}\ln\frac{R_1}{R_1-t}, \label{eq:Keff}
\end{equation}
where the second term represents the demagnetization energy density,
and the third term accounts for the exchange energy density associated with the radial domain state \cite{Sun_2014},
which is specific to tubular structures.
Figure \ref{fig:stable_states} shows the phase diagram of magnetic domain structures in the PMA nanotube,
validated through  \textmu M simulations (see Appendix \ref{sec:micromag}).
When $K_{{\rm eff}}>0$, the easy axes of magnetization are oriented radially;
otherwise, the easy axis aligns along the axial direction.

\section{Characteristics of domain walls}

In this section, we derive the energy densities per nanotube cross-sectional area
and present expressions for the effective fields acting on the DWs in PMA nanotubes.

Hereafter, we discuss structures where the thickness $t$ of the magnetic layer in the nanotube is smaller than the exchange length $\lambda=\sqrt{A/K_{{\rm eff}}}$.
The lack of edges in the nanotube's cross-section, which is perpendicular to the nanotube's axis, leads to high symmetry,
making it reasonable to assume uniform magnetization across the cross-section when $t<\lambda$.
Therefore, the magnetization structure at $z=X$ with the magnetization angle of $\phi$ is expressed as
$m_{r} =-Q\,{\tanh\frac{z-X}{\lambda}}$,
$m_{\varphi} ={\rm sech}{\frac{z-X}{\lambda}\sin\phi}$, and
$m_{z} ={\rm sech}{\frac{z-X}{\lambda}}\cos\phi $,
where the DW half-width is approximately equal to the exchange length $\lambda$ and  $Q$ represents the DW configuration, defined as $Q=+1$ for an up-down ($\uparrow\downarrow$)
wall (see Fig. \ref{fig:model}) and $Q=-1$ for a down-up ($\downarrow\uparrow$) wall \cite{Malozemoff_1979, Tatara_2008, Thiaville_2012, Martinez_2014, Rizinggard_2017}.

The PMA, DMI, and exchange energy densities per nanotube cross-sectional area are expressed as $w_{{\rm PMA,DMI,exc}}=\int{\rm d}z\,\varepsilon_{{\rm PMA,DMI,exc}}$.
The integral result of the PMA term does not depend on $\phi$ and thus does not affect the DW dynamics (see Eq. (\ref{eq:1DMeq_Gamma2})).

The integral result of the DMI term is given by 
\begin{equation}
w_{{\rm DMI}}\left(\phi\right)  =-\pi\mu_0 M_{{\rm s}}\lambda  H_{{\rm DMI}}Q\cos\phi, \\ \label{eq:w_DMI}
\end{equation}
where $H_{{\rm DMI}}=\frac{D}{\mu_0 M_{{\rm s}}\lambda}$ is the material-induced DMI field \cite{Rohart_2013, Martinez_2014, Soucaille_2016}.

The integral result of the $\phi$-dependent term of exchange interaction is given by 
\begin{equation}
w_{{\rm exc}}\left(\phi\right)=\mu_0 M_{{\rm s}} \lambda H_{{\rm a,exc}}^{{\rm tube}}\cos^{2}\phi, \\ \label{eq:w_exc}
\end{equation}
where $H_{{\rm a,exc}}^{{\rm tube}}=-\frac{4A}{\mu_0 M_{{\rm s}}}\frac{1}{R_{1}^{2}-R_{0}^{2}}\ln\frac{R_{1}}{R_{0}}$ is the curvilinear-geometry-induced anisotropy-like field \cite{Sheka_2015, Gaididei_2017}.
This result is obtained by integrating $\varepsilon_{{\rm exc}}=\frac{A}{r^{2}}\left(1+\frac{r^{2}}{\lambda^{2}}{\rm sech}^{2}\frac{z-X}{\lambda}-{\rm sech}^{2}\frac{z-X}{\lambda}\cos^{2}\phi\right)$
with respect to $X$ \cite{Htube_a_exc},
and is consistent with the result for IMA tubes with  an azimuthal easy axis of magnetization \cite{Hurst_2021}.
In the case where
$H_{{\rm a,exc}}^{{\rm tube}}$ is positive (negative), the Bloch
(N\'{e}el)-type DW is favored. 

\begin{figure*}[htbp]
\centering{}\includegraphics[scale=0.65]{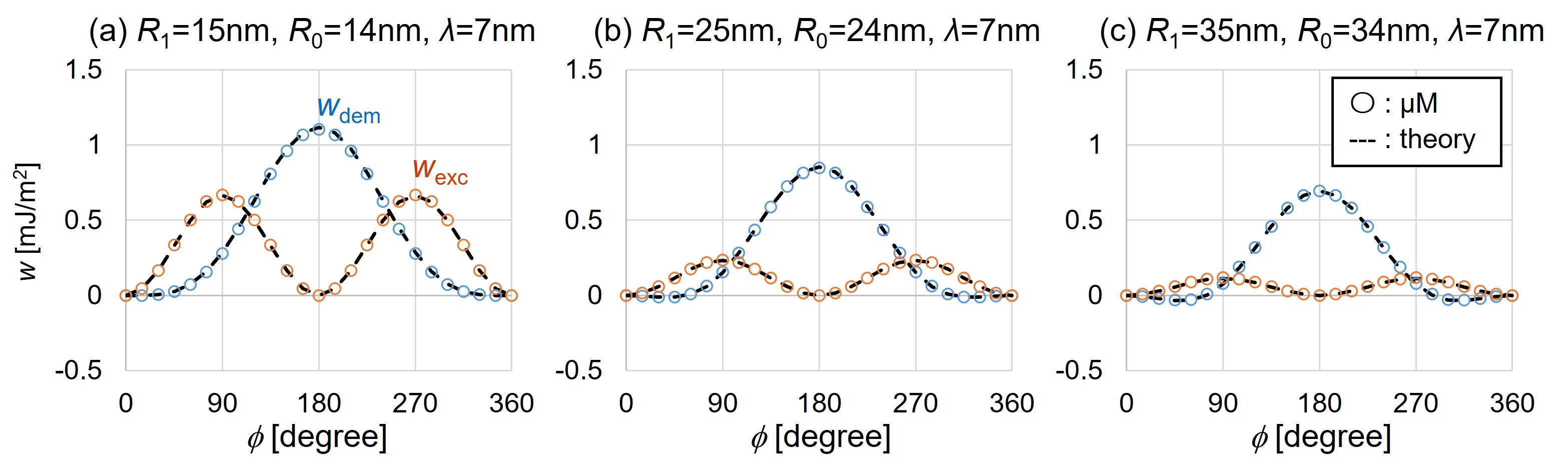}\caption{\protect\label{fig:w_compared}
Energy densities per cross-sectional area in PMA nanotubes as a function of magnetization angle.
The following parameters are used;
$A=10^{-11}\,{\rm J/m}$ and  $K_{\rm u}=0.83\times 10^{6}\,{\rm J/m^3}$.}
\end{figure*}

The energy density of the $\phi$-dependent term of magnetostatic interaction is given by the following equation, as derived in Appendix \ref{sec:derivation_demag}:
\begin{equation}
w_{{\rm dem}}\left(\phi\right)=\mu_0 M_{{\rm s}}\lambda\left(H_{{\rm a,dem}}^{{\rm {\rm tube}}}\cos^{2}\phi-\pi Q H_{{\rm D,dem}}^{{\rm tube}}\cos\phi\right),\label{eq:w_dem}
\end{equation}
where $H_{{\rm a,dem}}^{{\rm {\rm tube}}}$ and $H_{{\rm D,dem}}^{{\rm tube}}$
represent the anisotropy-like  demagnetizing field within the DW
\cite{DeJong_2015}
and DMI-like field which is specific to radially magnetized tubes
\cite{Sheka_2020}, respectively.
These fields are expressed as
\begin{align}
H_{{\rm a,dem}}^{{\rm {\rm tube}}}&=
\frac{2M_{{\rm s}}}{\rho_{1}^{2}-\rho_{0}^{2}}\int_{0}^{+\infty}{\rm d}y \, \frac{1}{\sinh y}\left( \frac{y}{\tanh y} - 1 \right) \nonumber \\ 
&\quad\times \Bigl(\rho_{1}^{2}I_{11}^{-1}\left(\rho_{1},\rho_{1},y\right)-2\rho_{1}\rho_{0}I_{11}^{-1}\left(\rho_{1},\rho_{0},y\right) \nonumber \\
&\quad\quad+\rho_{0}^{2}I_{11}^{-1}\left(\rho_{0},\rho_{0},y\right)\Bigr)
,\label{eq:Htube_a_dem}\\
H_{{\rm D,dem}}^{{\rm tube}}&=
\frac{2M_{{\rm s}}}{\rho_{1}^{2}-\rho_{0}^{2}}\int_{0}^{+\infty}{\rm d}y\, \tanh \frac{y}{2}  \int_{\rho_{0}}^{\rho_{1}}{\rm d}\rho\,\rho \nonumber \\
&\quad \times \left(\rho_{1}I_{11}^{1}\left(\rho_{1},\rho,y\right)-\rho_{0}I_{11}^{1}\left(\rho_{0},\rho,y\right)\right),\label{eq:Htube_D_dem}
\end{align}
where $\rho_{0,1}=R_{0,1}/\lambda$, $I_{\alpha\beta}^{\mu}\left(a,b,s\right)=\int_{0}^{\infty}{\rm d}x\,x^{\mu}J_{\alpha}\left(ax\right)J_{\beta}\left(bx\right){\rm e}^{-sx}$ and $J_{n}\left(x\right)$ is the $n$-th order Bessel function
of the first kind. 
The integrals are provided more specifically as follows \cite{Kausel_2012}:
$I_{11}^{1}\left(a,b,s\right) =  \frac{s\kappa}{2\pi ab\sqrt{ab}}\left(\frac{1-\frac{\kappa^{2}}{2}}{1-\kappa^{2}}E\left(\kappa\right)-K\left(\kappa\right)\right)$
and
$I_{11}^{-1}\left(a,b,s\right)= \frac{s}{\pi\sqrt{ab}}\left[\frac{1}{\kappa}E\left(\kappa\right)-\frac{\kappa
\left(a^{2}+b^{2}+\frac{s^{2}}{2} \right)}{2ab}K\left(\kappa\right)\right]
 +\frac{1}{4ab}\left[a^{2}+b^{2}+\left(a^{2}-b^{2}\right){\rm sgn}\left(a-b\right)\left(\Lambda\left(\nu,\kappa\right)-1\right)\right]$, 
where
$E\left(k\right)=\,\int_{0}^{\pi/2}{\rm d}\theta\sqrt{1-k^{2}\sin^{2}\theta}$ is the first elliptic integral function,
$K\left(k\right)=\,\int_{0}^{\pi/2}{\rm d}\theta\left(\sqrt{1-k^{2}\sin^{2}\theta}\right)^{-1}$ is the second elliptic integral function,
$\Pi\left(n,k\right)=\,\int_{0}^{\pi/2}{\rm d}\theta\left[\left(1-n\sin^{2}\theta\right)\sqrt{1-k^{2}\sin^{2}\theta}\right]^{-1}$
is the third  elliptic integral function,
$\kappa\left(a,b,s\right)=2\sqrt{\frac{ab}{\left(a+b\right)^{2}+s^{2}}}$,
$\nu\left(a,b\right)=\frac{4ab}{\left(a^{2}+b^{2}\right)}$,
and $\Lambda\left(n,k\right)=\frac{2}{\pi}\sqrt{\left(1-n\right)\left(n-k^{2}\right)/n}\,\Pi\left(n,k\right)$.
As shown in Fig. \ref{fig:w_compared},
the results of energy densities obtained from our \textmu M simulations closely align with those derived from the equations
presented in Eqs. (\ref{eq:w_exc}), (\ref{eq:w_dem}),
(\ref{eq:Htube_a_dem}), and (\ref{eq:Htube_D_dem}).

\begin{figure}[htbp]
\centering{}\includegraphics[scale=0.5]{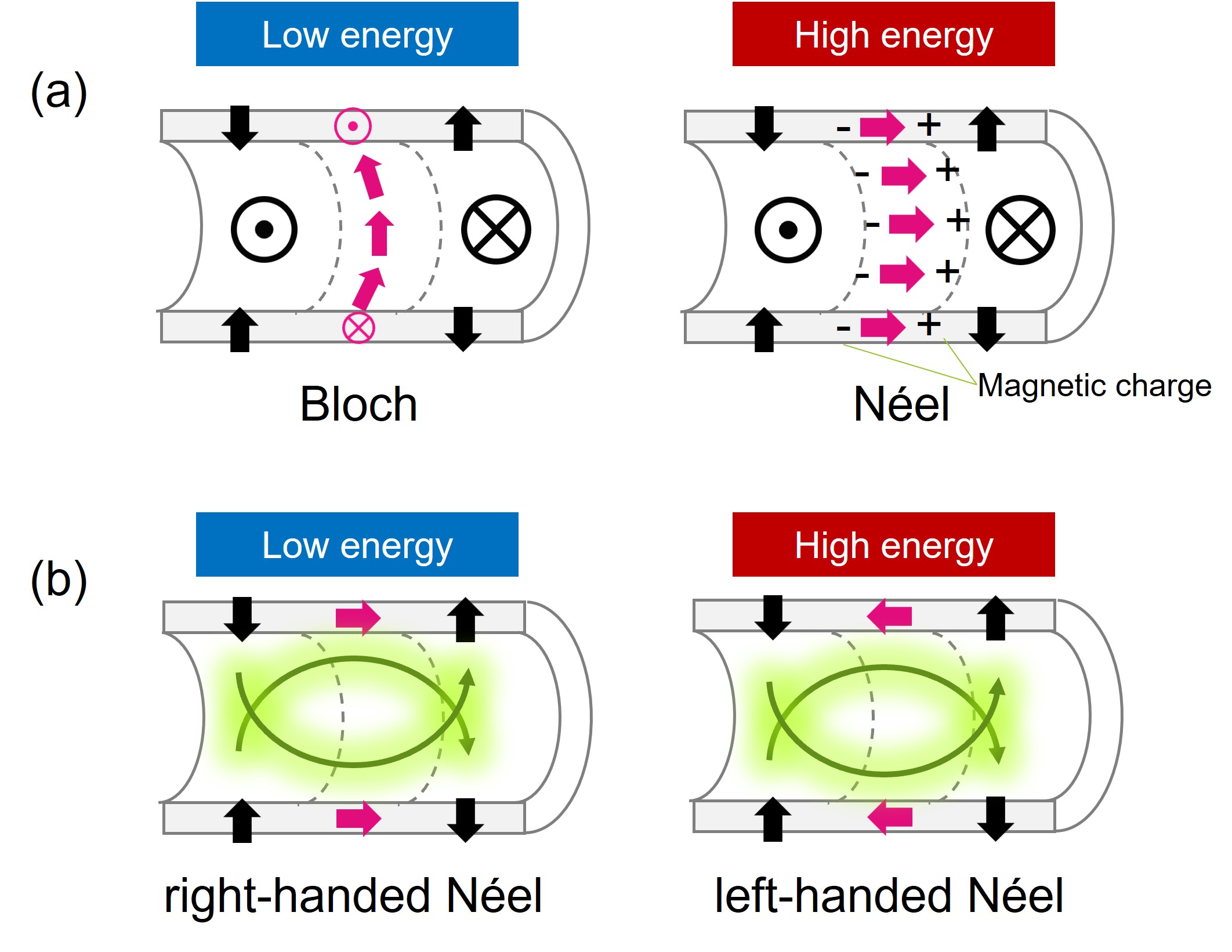}\caption{\protect\label{fig:demag}
Schematic illustration of geometrical effects arising from magnetostatic interactions in PMA tubes containing a DW.}
\end{figure}

The underlying physical mechanisms of the two terms in Eq. (\ref{eq:w_dem}) are shown in Fig. \ref{fig:demag}.
The first term of Eq. (\ref{eq:w_dem}) arises from the magnetic charges in the DW (Fig. \ref{fig:demag}(a)).
From the perspective of the unified theoretical framework \cite{Sheka_2020},
this phenomenon can be classified as the interactions between geometrical magnetostatic charges.
Since the DW in the radially magnetized tube does not generate a magnetic charge in the perfect Bloch-type DW ($\phi=\pm\pi/2$),
these states are minimal for the term.

The second term of Eq. (\ref{eq:w_dem}) arises from the stray field from the magnetic domains in the radially magnetized tube (Fig. \ref{fig:demag}(b)).
When the DW magnetization is aligned with the stray field,  
corresponding to the right-handed N\'{e}el-type DW, the energy of this configuration is at a minimum.
This term consists of not only the interaction between geometrical magnetostatic charges
but also the interaction between geometrical and surface magnetostatic charges \cite{Sheka_2020}.
The fundamental basis of this field lies in the asymmetry in the distribution of magnetic charges on the inner and outer surfaces of the nanotube.
It should be noted that this field is specific to radially magnetized tubes and is not present in flat ribbons and other tubes with a different easy axis of magnetization.

\begin{figure*}[htbp]
\centering{}\includegraphics[scale=0.55]{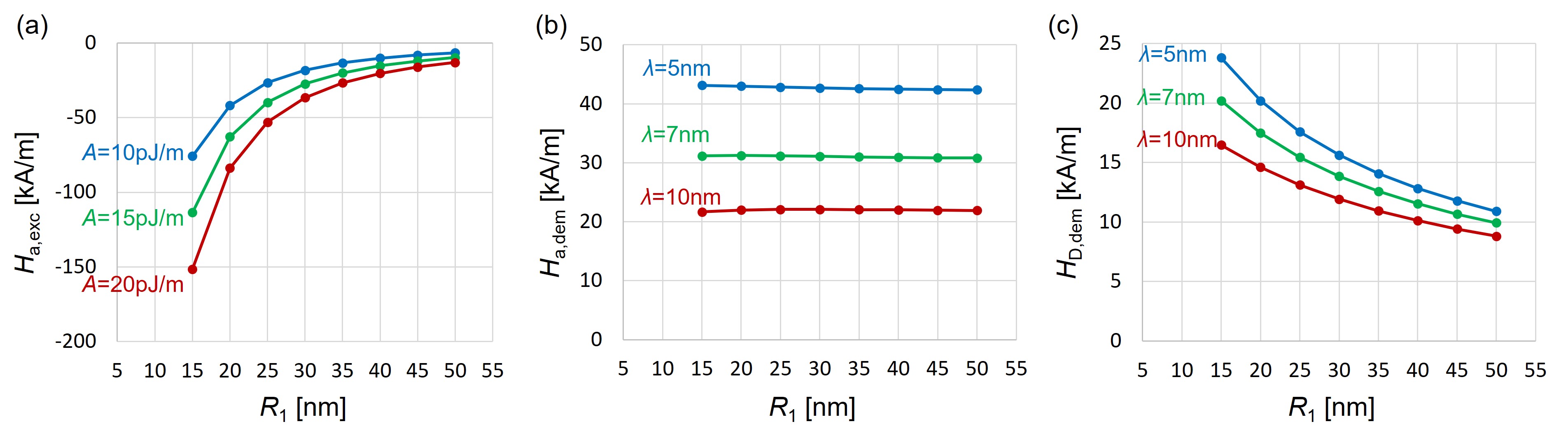}\caption{\protect\label{fig:fields}
Radius dependence of the effective field acting on the DW in PMA nanotubes with $t=1\,{\rm nm}$.}
\end{figure*}

Figure \ref{fig:fields} shows the radius dependence of the each effective field acting on the DW in the PMA nanotube,
as evaluated using the derived equations. 
The absolute values of the effective fields $H_{{\rm a,exc}}^{{\rm tube}}$
and $H_{{\rm D,dem}}^{{\rm tube}}$ increase as the tube radius decreases,
while $H_{{\rm a,dem}}^{{\rm tube}}$ remains approximately constant regardless of the radius.

The exchange interaction increases with greater curvature,
leading to a corresponding increase in  $\left|H_{{\rm a,exc}}^{{\rm tube}}\right|$ as the radius decreases,
shown in Fig. \ref{fig:fields}(a).
Since this field is negative, N\'{e}el-type DWs are favored in systems with smaller radii.
This characteristic is similar to the behavior observed in flat nanoribbons, where narrower wire width favor N\'{e}el-type DWs
\cite{DeJong_2015}. However, unlike in flat nanoribbons where this characteristic is attributed to magnetostatic interactions,
in nanotubes, it is due to exchange  interactions, indicating that the origins of the two characteristics are different.

The anisotropy-like  demagnetizing field from the magnetic charges in the DW,
$H_{{\rm a,dem}}^{{\rm tube}}$,
is primarily determined by the DW width and is thus independent of the tube radius, as shown in Fig. \ref{fig:fields}(b).
This characteristic differs from the behavior observed in flat nanoribbons, where the anisotropy-like  demagnetizing field depends on the wire width
\cite{DeJong_2015}. In flat nanoribbons, as the wire width decreases, this effective field decreases and eventually reverses its sign at a certain width, favoring N\'{e}el-type DWs over Bloch-type DWs.
This discrepancy arises from the presence of edges in nanowires.
The lack of edges in the tube's cross-section prevents the emergence of magnetic charges within Bloch-type DWs,
making them preferred ($H_{{\rm a,dem}}^{{\rm tube}}>0$).

The stray field generated by the magnetic domains, $H_{{\rm D,dem}}^{{\rm tube}}$, decreases as the radius increases, as shown in Fig. \ref{fig:fields}(c).
This is because the distance between opposing magnetic domains
located on either side of the tube's central axis increases. 
Furthermore, a reduction in the DW width corresponds to the expansion of the magnetic domain regions, 
thereby enhancing the stray field.

\begin{figure}[htbp]
\centering{}\includegraphics[scale=0.4]{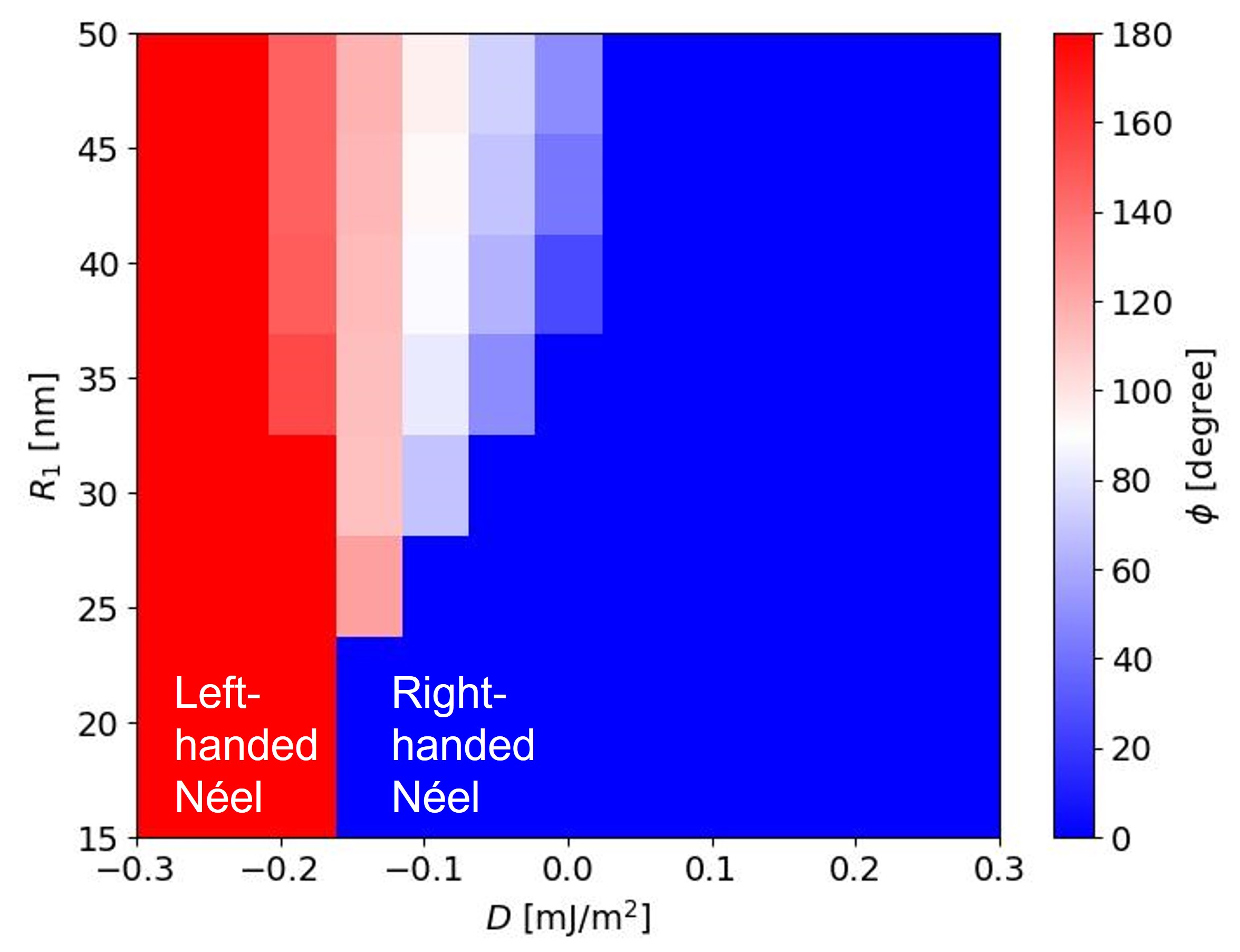}\caption{\label{fig:angle}
Magnetization angle of DW in equilibrium states of  PMA nanotubes.
The following parameters are used;
$A=10^{-11}\,{\rm J/m}$, $K_{\rm u}=0.83\times 10^{6}\,{\rm J/m^3}$, and $t=1\,{\rm nm}$.}
\end{figure}

In PMA nanotubes, the total anisotropy and DMI fields acting on a DW
are written by $H_{{\rm a}} =H_{{\rm a,exc}}^{{\rm tube}}+H_{{\rm a,dem}}^{{\rm tube}}$ and
$H_{{\rm D}}=H_{{\rm D,dem}}^{{\rm tube}}+H_{{\rm DMI}}$, respectively.
The magnetization angle of the DW in the equilibrium
states can be expressed by the following equation \cite{Je_2013}:
\begin{equation}
\begin{aligned}
\phi_{{\rm eq}}=\begin{cases}
0 &
\frac{2H_{{\rm a}}}{\left|\pi H_{{\rm D}}\right|}<1
\cap H_{{\rm D}}>0\\
\pi &
\frac{2H_{{\rm a}}}{\left|\pi H_{{\rm D}}\right|}<1
\cap H_{{\rm D}}<0\\
{\rm arccos}\left(\frac{\pi H_{{\rm D}}}{2H_{{\rm a}}}\right) &
\frac{2H_{{\rm a}}}{\left|\pi H_{{\rm D}}\right|}>1
\end{cases},\label{eq:phi_eq}
\end{aligned}
\end{equation}
where we assume that $Q=+1$ and consider only the range of $\phi$ from 0 to $\pi$ due to the symmetry.
The equilibrium angle is influenced by both the material-induced DMI constant and the radius of the nanotube, as shown in Fig. \ref{fig:angle}.
As the radius decreases, the intermediate region between the Bloch- and N\'{e}el-type DWs contracts,
reflecting the radius dependence of $H_{{\rm a,exc}}^{{\rm tube}}$.
Additionally, the boundary between the left- and right-handed N\'{e}el configurations shift toward the negative side of $D$ as the radius decreases,
reflecting the radius dependence of $H_{{\rm D,dem}}^{{\rm tube}}$.
For $R_{1}\lesssim 25\, {\rm nm}$, no intermediate states exist between the Bloch- and  N\'{e}el-type DWs,
leading to a discontinuous transition between left- and right-handed N\'{e}el configurations.
This discontinuity is due to the absence of an effective field to stabilize a Bloch-type DW,
as $H_{\rm a}$ becomes negative for $R_{1}\lesssim 25\,{\rm nm}$.
This corresponds to the condition $2H_{\rm a}/\left|\pi H_{\rm D}\right| < 1$ in Eq. (8).
In the PMA nanotube composed of typical 3$d$ transition metal alloys (with $M_{\rm s}= 10^{6}\,{\rm A/m}$ assumed in this model),
it is confirmed that geometrical effects contribute on the order of $0.1\,{\rm mJ/m^2}$,
comparable to the material-induced DMI \cite{Torrejon_2014, Soucaille_2016}.

In flat nanoribbons,
$H_{{\rm a}}$ arises exclusively from the anisotropy-like demagnetizing field arising from magnetic charges within the DW,
while $H_{{\rm D}}$ does not include any geometry-dependent terms.
Consequently, in flat nanoribbons,
the total anisotropy field is influenced by the nanoribbon width \cite{DeJong_2015},
whereas the total DMI field remains invariant to changes in dimensions.

\section{Spin-orbit torque driven domain wall motion}

In this section, we present the numerical results of current-induced DW motion in PMA nanotubes under the influence of STT and SOT.

The dynamics of DWs are described by the following differential equations within the framework of an analytical 1DM
\cite{Malozemoff_1979, Tatara_2008, Thiaville_2012, Martinez_2014, Rizinggard_2017}:
\begin{equation}
\begin{aligned}
\frac{{\rm d}\phi}{{\rm d}t} & =\frac{\gamma_0}{1+\alpha^{2}}\left(-Q\Gamma_{1}-\alpha\Gamma_{2}\right),\\
\frac{{\rm d}\left(X/\lambda\right)}{{\rm d}t} & =\frac{\gamma_0}{1+\alpha^{2}}\left(-\alpha\Gamma_{1}+Q\Gamma_{2}\right),\\
\Gamma_{1}  &=Q\frac{\pi}{2}\xi_{{\rm DLT}}j\cos\phi,\\
\Gamma_{2}  &=Q\xi_{{\rm STT}}j+\frac{1}{2\mu_0 M_{{\rm s}}\lambda}\frac{\partial w}{\partial\phi},\label{eq:1DMeq_Gamma2}
\end{aligned}
\end{equation}
where $\gamma_0$ is the gyromagnetic ratio, $\alpha$ is the Gilbert
damping, $j$ is the current density, and
$w=w_{{\rm DMI}}\left(\phi\right)+w_{{\rm exc}}\left(\phi\right)+w_{{\rm dem}}\left(\phi\right)$ is the energy density per
cross-sectional area of the nanotube.
The damping-like torque (DLT) efficiency induced by spin Hall effects is given by
$\xi_{{\rm DLT}} =\frac{\hbar\theta_{{\rm SH}}}{2 \mu_0 M_{{\rm s}}\left|e\right|t}$,
and the STT efficiency is defined as
$\xi_{{\rm STT}} =-\frac{\hbar P}{2 \mu_0 M_{{\rm s}}\left|e\right|\lambda}$.
Here,
$e$ is the elementary charge, $\hbar$ is the Dirac constant,
$\theta_{{\rm SH}}$ is the spin Hall angle (SHA)
, and $P$ is the spin polarization.
We assume that the same current density is present in both the HM and FM layers \cite{Umetsu_2025}.
Based on the findings from the previous section,
$\Gamma_{2}$ in Eq. (\ref{eq:1DMeq_Gamma2}) can be expressed as
\begin{equation}
\begin{aligned}
\Gamma_{2}=&\,\xi_{{\rm STT}}j-\frac{1}{2}\left(
H_{{\rm a,exc}}^{{\rm tube}}+H_{{\rm a,dem}}^{{\rm tube}}\right)\sin2\phi \\ 
&+Q\frac{\pi}{2}\left(
H_{{\rm D,dem}}^{{\rm tube}}+H_{{\rm DMI}}\right)\cos\phi.
\end{aligned}
\end{equation}

We assume that the HM layer consists of materials with a negative SHA, such as W, Ta, or Hf \cite{Emori_2013, Torrejon_2014, Fritz_2018, Umetsu_2025}.
In the case where the FM layer is composed of CoFeB,
it is known that the sign of DMI is positive for a W layer \cite{Kim_2018, Umetsu_2025},
negative for a Hf layer \cite{Torrejon_2014}, and can take both signs for a Ta layer \cite{Emori_2013, Torrejon_2014, Conte_2015}.
Consequently, we carried out simulations for different values of DMI: positive, negative, and zero.

We disregard the field-like torque (FLT) as it does not serve as a driving force for DW motion in PMA layers.
The FLT acts on the DW as an effective field directed along the azimuthal axis in PMA nanotubes.
For example, it has been reported that in HM/CoFeB/MgO structures with HM layers composed W or Ta,
the FLT is approximately one order of magnitude smaller than the DLT \cite{Lau_2017}.
Although the contribution of the {\O}rsted field is included in the FLT,
for the parameters of interest (specifically, $R_1=O({10\, \rm nm})$ and $t=O({1\, \rm nm})$),
this contribution is more than an order of magnitude smaller than the DLT \cite{Oersted}.

\begin{figure*}[htbp]
\centering{}\includegraphics[scale=0.45]{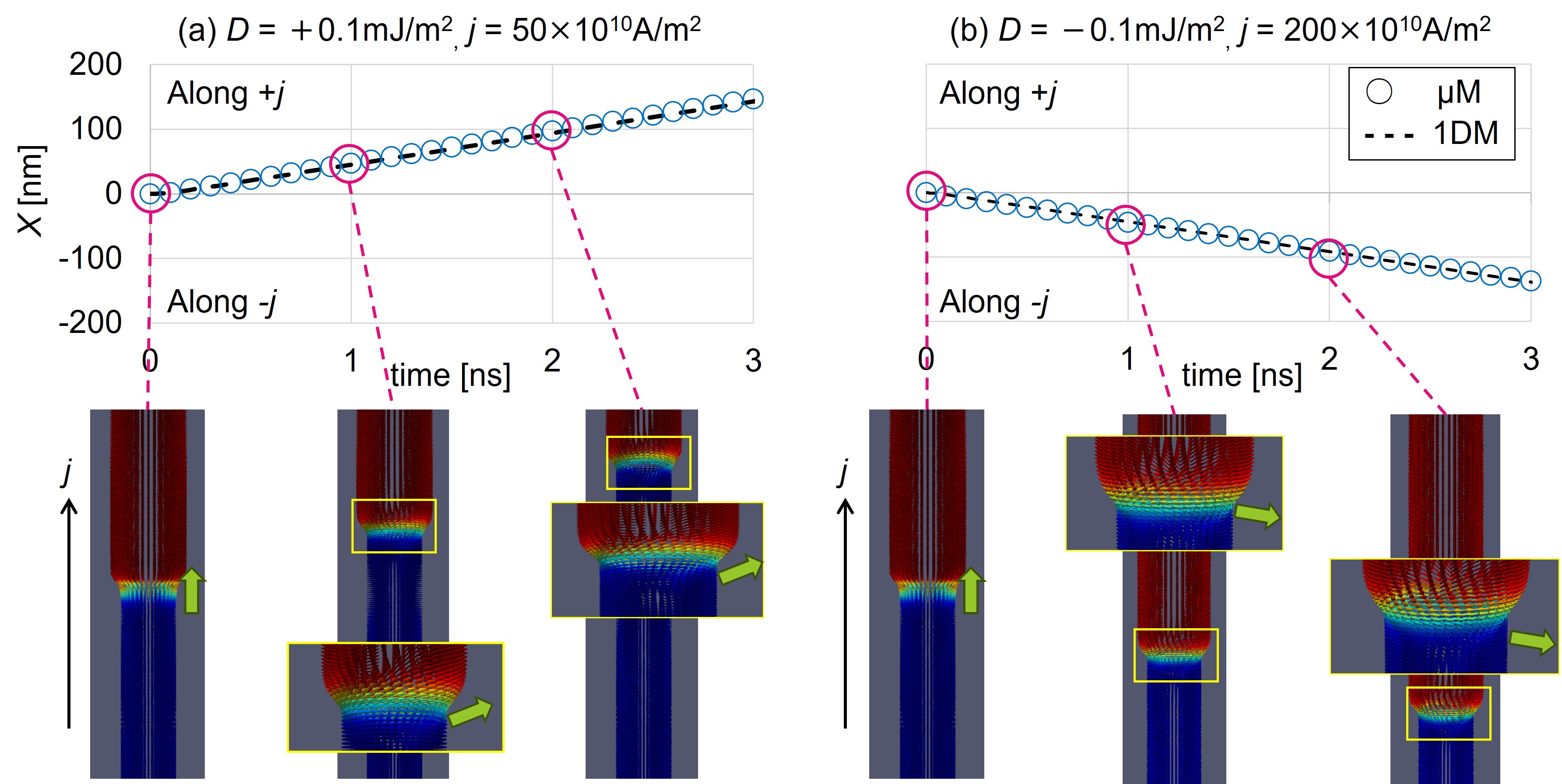}\caption{\protect\label{fig:dynamics}
Transient position of DW in the PMA nanotube under applied currents.
Images are snapshots of the magnetization structure by
\textmu M simulations.
The enlarged views show the magnetization structure in the vicinity of the DW, with the green arrows indicating the direction of magnetization of the DW.
The following parameters are used;
$A=10^{-11}\,{\rm J/m}$, $K_{\rm u}=0.83\times 10^{6}\,{\rm J/m^3}$, $\alpha=0.2$, $P=0.5$, $\theta_{{\rm SH}}=-0.2$, $R_1=25\,{\rm nm}$, and $t=1\,{\rm nm}$.}
\end{figure*}

Figure \ref{fig:dynamics} shows the transient position of the DW in the PMA nanotube under applied currents.
The results from 1DM and \textmu M simulations are consistent,
demonstrating the validity of the approximation that the DW width is constant in the 1DM.
The magnetization angle of the DW remains constant during current flow,
and the direction of DW motion depends on the sign of $D$.
The DWs driven by SOT are slightly displaced from the perfect Bloch-type DW ($\phi=\pi/2$).
In systems with a negative SHA,
it has been theoretically shown that the DW moves
in the direction of current (electron) flow for the right (left)-handed N\'{e}el-type DW
\cite{Kim_2022, Umetsu_2025}.
Our computational results are consistent with these predictions.
These results confirm that the SOT-driven motion in PMA nanotubes is qualitatively similar to that observed in
PMA flat nanoribbons.

\begin{figure*}[htbp]
\centering{}\includegraphics[scale=0.6]{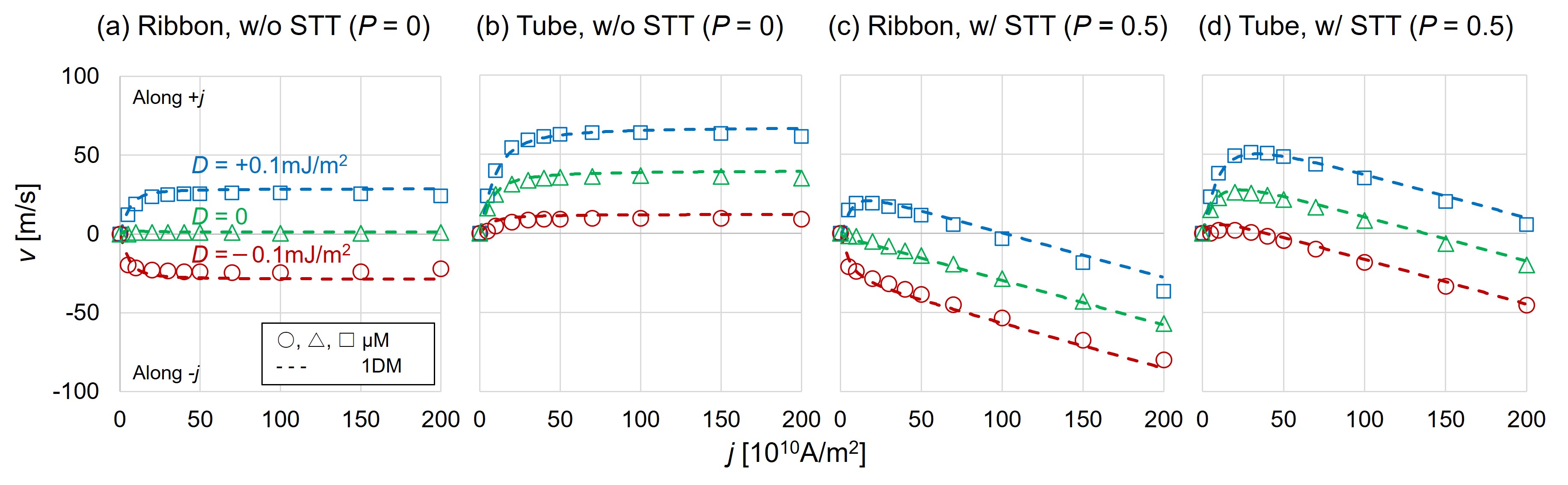}\caption{\protect\label{fig:v-j_1}
DW velocity as a function of current density ($v$-$j$ curve) in the PMA flat nanoribbon with a width of $50\,{\rm nm}$ and the PMA nanotube with $R_1=25\,{\rm nm}$.
The anisotropy-like field ($H_{\rm a}$) used for the flat nanoribbons in the 1D simulation was derived from energy calculations obtained through \textmu M simulations.
The pulse widths for the calculations are 2$\,$ns.
The following parameters are used;
$A=10^{-11}\,{\rm J/m}$, $K_{\rm u}=0.83\times 10^{6}\,{\rm J/m^3}$, $\alpha=0.2$, $\theta_{{\rm SH}}=-0.2$, and  $t=1\,{\rm nm}$.}
\end{figure*}

The impact of geometrical effects on SOT-driven DW motion in PMA nanotubes is clarified by comparing it to
the DW velocity as a function of current density ($v$-$j$ curve) in PMA flat nanoribbons, as shown in Fig. \ref{fig:v-j_1}.
In systems where only SOT is present and STT is absent,
the direction of DW motion in the flat nanoribbon reverses with a change in the sign of $D$, but the speeds of DW motion are the same (Fig. \ref{fig:v-j_1}(a)).
In contrast, in the nanotube, the direction of DW motion does not necessarily reverse with a change in the sign of $D$,
and the speeds of DW motion vary accordingly (Fig. \ref{fig:v-j_1}(b)).
It should be noted that a geometry-induced positive DMI contribution is always present in our PMA nanotube models.

In systems where both STT and SOT are present, SOT serves as the driving force for DW motion,
while STT indirectly affects DW motion by modulating the DW magnetization angle \cite{Umetsu_2025}.
In the case of $D=+0.1\,{\rm mJ/m^{2}}$,
the higher DW speed observed in the nanotube (Fig. \ref{fig:v-j_1}(d)) compared to the flat nanoribbon (Fig. \ref{fig:v-j_1}(c))
in the low current density region.
This is attributed to the presence of a specific positive-sign DMI-like field within the PMA nanotube.
The decrease in DW velocity with increasing current density in the high current region
is the characteristic of systems with a negative SHA and a positive DMI \cite{Umetsu_2025}.
This behavior reflects the transition of the DW from a right-handed N\'{e}el-type DW to a left-handed N\'{e}el-type DW,
driven by increasing STT with higher currents.

In PMA nanotubes, a geometry-induced positive DMI-like field enables DW motion in the direction of the current flow even when $D\leq 0$ and $\theta_{\rm SH}<0$,
which differs from PMA flat nanoribbons \cite{Kim_2022}.
As shown in Fig. \ref{fig:v-j_1}(c),
the DW in the PMA flat nanoribbon only moves in the direction of electron flow when $D\leq 0$,
whereas in the PMA nanotube,
the DW moves in the direction of the current flow in the low current density region even when $D\leq 0$ shown in Fig. \ref{fig:v-j_1}(d).
The behavior observed in the PMA nanotube is consistent with the total DMI field,
where $H_{{\rm D}}=H_{{\rm D,dem}}^{{\rm tube}}+H_{{\rm DMI}}>0$.

\begin{figure}[htbp]
\centering{}\includegraphics[scale=0.6]{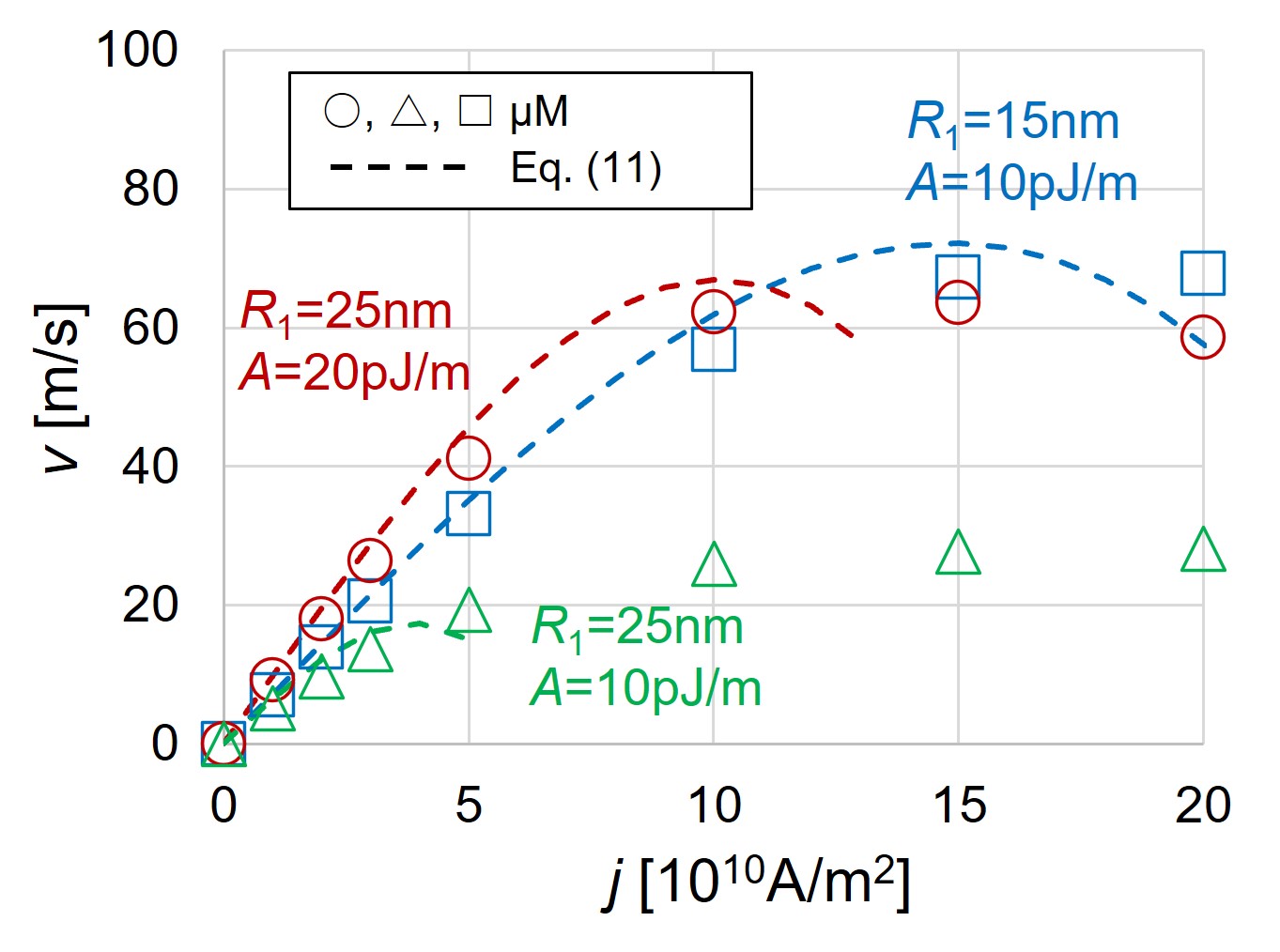}\caption{\protect\label{fig:v-j_2}
Comparison of $v$-$j$ curves for PMA nanotubes with different radius $(R_1)$ and exchange stiffness constant $(A)$.
The pulse widths for the calculations are 5$\,$ns.
The following parameters are used;
 $K_{\rm u}=0.83\times 10^{6}\,{\rm J/m^3}$, $D=0$, $\alpha=0.2$, $P=0.5$, $\theta_{{\rm SH}}=-0.2$, and  $t=1\,{\rm nm}$.}
\end{figure}

Geometrical effects in PMA nanotubes can stabilize a N\'{e}el-type DW structure during current flow, even when the material-induced DMI is weak, enabling highly efficient SOT-driven DW motion. 
Figure \ref{fig:v-j_2} shows that structures with smaller radii and materials with larger exchange stiffness exhibit higher SOT efficiency.
The reduction of radius and the increase in exchange stiffness promote the stabilization of N\'{e}el-type DWs,
as they increase the exchange energy associated with the Bloch-type DWs (which corresponds to the decrease in $H_{{\rm a,exc}}^{{\rm tube}}$ with reduced $R_1$ and increased $A$, as shown in Fig. \ref{fig:fields}(a)).
Additionally, the reduction in radius enhances the stray field from the magnetic domains acting on the DW
(which corresponds to the increase in  $H_{{\rm D,dem}}^{{\rm tube}}$ with reduced $R_1$, as shown in Fig. \ref{fig:fields}(c)), further stabilizing right-handed N\'{e}el-type DWs.
These combined effects suggest that tailoring geometric parameters can effectively induce and stabilize N\'{e}el-type DWs, thereby achieving higher-speed SOT-driven DW motion without relying solely on the material-induced DMI.
Furthermore, to achieve the same speed of DW motion, a lower current density is required in PMA nanotubes compared to PMA flat nanoribbons.

The velocity  of DW  motion induced by the SOT for the N\'{e}el-type DW in the low current regime is described by the following equation \cite{Rizinggard_2017}:
\begin{equation}
v = -\frac{\gamma_0\lambda}{\alpha}\frac{\pi}{2}\xi_{\rm DLT}j
\left[1-
\frac{\left(\alpha^{-1}\frac{\pi}{2}\xi_{\rm DLT}+\xi_{\rm STT}\right)^2}{2\left(\frac{\pi}{2}H_{\rm D}-H_{\rm a}\right)^2}j^2\right].
\end{equation}
The second term reflects the deceleration that occurs
when the DW deviates from the perfect N\'{e}el-type DW ($\phi=0,\pi$) during current flow. 
As the radius of the nanotube decreases or the exchange stiffness increases,
$H_{{\rm a}} =H_{{\rm a,exc}}^{{\rm tube}}+H_{{\rm a,dem}}^{{\rm tube}}$ increases in the negative direction,
while $H_{{\rm D}}=H_{{\rm D,dem}}^{{\rm tube}}+H_{{\rm DMI}}$ increases in the positive direction (see Fig. \ref{fig:fields}),
resulting in a reduction of the deceleration effect.
In flat nanoribbons, which represent the limiting case where the radius is infinitely large,
the deceleration effect is more pronounced compared to nanotubes, leading to slower DW motion.

\section{Conclusions}

In this study, we analytically investigated the characteristics of chiral DWs induced by geometrical effects in radially magnetized nanotubes with PMA.
We derived expressions for the effective magnetic fields acting on DWs and formulated the SOT-driven DW motion in PMA nanotubes using a one-dimensional theory.
This theory was validated through micromagnetic simulations.

Our findings demonstrate that the stray field originating from the magnetostatic interactions specific to PMA nanotubes can have a contribution comparable to the material-induced DMI for diameters in the sub-$100\,{\rm nm}$ range. 
Consequently, the direction and speed of DW motion in PMA nanotubes may differ from those observed in flat nanoribbons composed of the same material.

The DMI-like field resulting from magnetostatic interactions depends on the tube diameter, enabling the control of SOT-driven DW motion speeds through geometric adjustments rather than the material-induced DMI, which is sensitive to deposition conditions \cite{Quinsat_2017, Kim_2018, Umetsu_2025}.
Additionally, the anisotropy-like field induced by curvilinear geometry, arising from exchange interactions, 
helps to stabilize N\'{e}el-type DWs as tube diameters decrease or exchange stiffness constants of the magnetic layer increase.
Utilizing these effects is expected to result in more efficient SOT-driven DW motion.
These insights are anticipated to significantly expand the potential applications of PMA nanotubes.

\appendix

\section{Micromagnetic simulations \label{sec:micromag}}

\begin{figure*}[htbp]
\centering{}\includegraphics[scale=0.55]{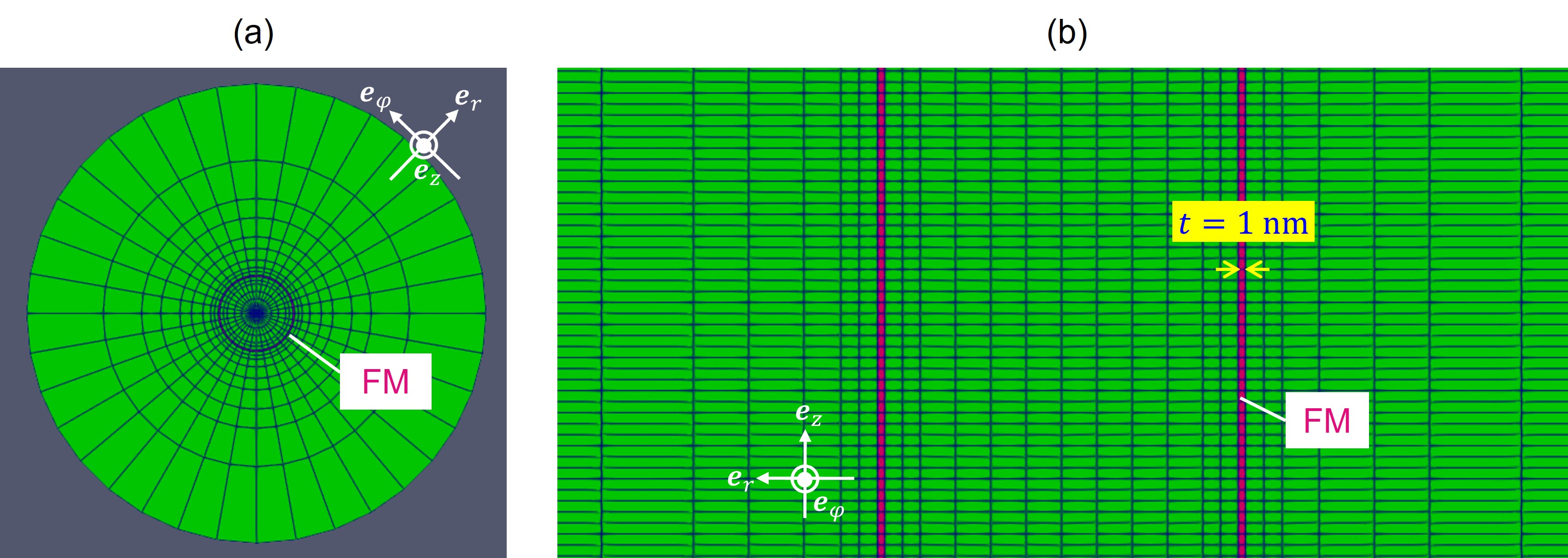}\caption{\protect\label{fig:mesh}
Top (a) and cross-sectional (b) views of the mesh structure in \textmu M simulations.
Green and pink represent the non-magnetic and magnetic regions, respectively.}
\end{figure*}

We conducted micromagnetic (\textmu M) simulations \cite{Martinez_2014} of a DW inside a PMA nanotube to validate our analytical formulation.
Our \textmu M program utilizes the finite volume method \cite{Rupp_2014},
and the computational domain includes both ferromagnetic and non-magnetic regions, as shown in Fig. \ref{fig:mesh}(a) and (b).
This method is compatible with unstructured meshes,
and we therefore employed uniform unstructured meshes in the azimuthal direction to minimize any extraneous effects arising from the mesh geometry.
The height of the nanotube in our \textmu M simulation model is $700\,{\rm nm}$ (within the range of $z = -350$ to $+350\,{\rm nm}$).
To minimize the impact of end effects on the DW,
we conducted shift simulations with pulse width conditions designed to keep the DW within the range of $z = -200$ to $+200\,{\rm nm}$.

\section{Derivation of magnetostatic interaction terms\label{sec:derivation_demag}}

In our PMA nanotube model, the surface and volume magnetic charge densities are given by
\begin{align}
\frac{\sigma_{{\rm m}}\left(\bm{x}\right)}{M_{\rm s}} & =\begin{cases}
-Q\tanh\frac{z-X}{\lambda} & {\rm at}\,r=R_{1}\\
+Q\tanh\frac{z-X}{\lambda} & {\rm at}\,r=R_{0}\\
\end{cases}, \label{eq:sigma_m}\\
\frac{\rho_{\rm m}\left(\bm{x}\right)}{M_{\rm s}}   &=\left(\frac{Q}{r}+\frac{\cos\phi}{\lambda}{\rm sech}{\frac{z-X}{\lambda}}\right){\tanh\frac{z-X}{\lambda}}.\label{eq:rho_m}
\end{align}
From Eqs. (\ref{eq:Edem}), (\ref{eq:sigma_m}), and (\ref{eq:rho_m}), we obtain 
\begin{widetext}
\begin{align}
E_{{\rm dem,bb}}
= & \, \frac{\pi}{2} \mu_0 M_{{\rm s}}^{2}\lambda^{3}\int_{0}^{+\infty}{\rm d}\kappa\int_{\rho_{0}}^{\rho_{1}}{\rm d}\rho\,J_{0}\left(\rho\kappa\right)\rho\int_{\rho_{0}}^{\rho_{1}}{\rm d}\rho'J_{0}\left(\rho'\kappa\right)\rho'
\int_{-\infty}^{+\infty}{\rm d}x\int_{-\infty}^{+\infty}{\rm d}y\nonumber \\
 & \times\left(\frac{Q}{\rho}+\cos\phi\,{\rm sech}x\right)\left(\frac{Q}{\rho'}+\cos\phi\,{\rm sech}y\right){\displaystyle \tanh x}\,{\displaystyle \tanh}y\,{\rm e}^{-\left|x-y\right|\kappa},\label{eq:Edem,bb}\\
E_{{\rm dem,bs}}
= & \,\pi Q\mu_0 M_{{\rm s}}^{2}\lambda^{3}\int_{0}^{+\infty}{\rm d}\kappa\int_{\rho_{0}}^{\rho_{1}}{\rm d}\rho\,J_{0}\left(\rho\kappa\right)\rho\left[J_{0}\left(\rho_{0}\kappa\right)\rho_{0}-J_{0}\left(\rho_{1}\kappa\right)\rho_{1}\right]\int_{-\infty}^{+\infty}{\rm d}x\int_{-\infty}^{+\infty}{\rm d}y\nonumber \\
 & \times\left(\frac{Q}{\rho}+\cos\phi\,{\rm sech}x\right){\displaystyle \tanh}x\,{\displaystyle \tanh}y\,{\rm e}^{-\left|x-y\right|\kappa},\label{eq:Edem,bs}
\end{align}
\end{widetext}
where  $J_{n}\left(x\right)$ is the $n$-th order Bessel function
of the first kind. The term $E_{{\rm dem,ss}}$, which is independent
of $\phi$, is not included in the calculation as it does not affect
the DW dynamics. Equations (\ref{eq:Edem,bb}) and (\ref{eq:Edem,bs})
include the terms proportional to $\cos\phi$, and Eq. (\ref{eq:Edem,bb}) has the term proportional to $\cos^{2}\phi$.
By dividing these by the tube cross-sectional area, the linear
and quadratic terms of $\cos\phi$ can be expressed as
\begin{widetext}
\begin{align}
w_{1}
= & \,Q\frac{\mu_0 M_{{\rm s}}^{2}\lambda}{\rho_{1}^{2}-\rho_{0}^{2}}\cos\phi\int_{0}^{+\infty}{\rm d}\kappa\left[\rho_{1}J_{1}\left(\rho_{1}\kappa\right)-\rho_{0}J_{1}\left(\rho_{0}\kappa\right)\right]\int_{\rho_{0}}^{\rho_{1}}{\rm d}\rho\,J_{1}\left(\rho\kappa\right)\rho
\int_{-\infty}^{+\infty}{\rm d}x\,{\displaystyle \tanh}x\int_{-\infty}^{+\infty}{\rm d}y\,\frac{{\tanh}y}{\cosh y}\,{\rm e}^{-\left|x-y\right|\kappa}, \label{eq:w1_2} \\
w_{2}
= & \,\frac{1}{2}\frac{\mu_0 M_{{\rm s}}^{2}\lambda}{\rho_{1}^{2}-\rho_{0}^{2}}\cos^{2}\phi\int_{0}^{+\infty}{\rm d}\kappa\left[\frac{\rho_{1}J_{1}\left(\rho_{1}\kappa\right)-\rho_{0}J_{1}\left(\rho_{0}\kappa\right)}{\kappa}\right]^{2}
\int_{-\infty}^{+\infty}{\rm d}x\,
\frac{{\tanh}x}{\cosh x}
\int_{-\infty}^{+\infty}{\rm d}y\,
\frac{{\tanh}y}{\cosh y}\,
{\rm e}^{-\left|x-y\right|\kappa}. \label{eq:w2_2}
\end{align}
\end{widetext}
Using the identity $\int_{-\infty}^{+\infty}{\rm d}x\,f(x) \int_{-\infty}^{+\infty}{\rm d}y\,\frac{{\tanh}y}{{\cosh}y}{\rm e}^{-\left|x-y\right|\kappa}=2\kappa \int_{0}^{+\infty}{\rm d}y \,{\rm e}^{-y \kappa} \int_{-\infty}^{+\infty}{\rm d}x\,f(x)/\cosh (y-x) $, which holds for the odd function $f(x)$,
and integrating both Eq. (\ref{eq:w1_2}) and Eq. (\ref{eq:w2_2}) with respect to $x$, we can compare the results to Eqs. (\ref{eq:w_DMI}) and (\ref{eq:w_exc}), respectively, to obtain Eqs. (\ref{eq:Htube_a_dem})  and (\ref{eq:Htube_D_dem}) .
\bibliography{reference}

\end{document}